\documentclass[10pt,a4paper]{article}
\usepackage[utf8]{inputenc}
\usepackage{amsmath}
\usepackage{booktabs}
\usepackage{caption}
\usepackage{amsfonts}
\usepackage{amssymb}
\usepackage{makeidx}
\usepackage{graphicx}
\usepackage{bm}
\usepackage{hyperref}
\usepackage[left=2cm,right=2cm,top=2cm,bottom=2cm]{geometry}

\usepackage{authblk}

\title{Deviations to cobimaximal mixing in a 3+1 neutrino scenario}

\author[1]{Diana C. Rivera-Agudelo \thanks{\href{mailto:diana.rivera11@usc.edu.co}
{\texttt{diana.rivera11@usc.edu.co}}}}

\author[1]{S. L. Tostado \thanks{\href{mailto:sergio.tostado00@usc.edu.co}
{\texttt{sergio.tostado00@usc.edu.co}}}}

\affil[1]{Universidad Santiago de Cali, Facultad de Ciencias Básicas, Campus
Pampalinda, Calle 5 No. 62-00, Código Postal 76001, Santiago de Cali,
Colombia.}

\date{}

\begin{document}
\maketitle

\begin{abstract}
Cobimaximal mixing remains a viable approximate pattern since the predicted atmospheric angle ($\pi/4$) and Dirac CP phase ($-\pi/2$) are compatible with current experimental data at the $3\sigma$ level. In this work, we investigate how an additional sterile neutrino state breaks the cobimaximal mixing and the underlying $\mu-\tau$ reflection symmetry in a $3+1$ neutrino scenario. Given the strong constraints already imposed on active-sterile mixings, we examine the resulting restrictions on the model parameter space, assuming a small symmetry breaking induced by the sterile state. We derive predictions for the relevant model parameters, active-sterile mixing angles, and Dirac CP-violating phases, by considering small deviations from the symmetric limit. We also analyze the allowed combinations of Majorana phases through their impact on the effective Majorana mass governing neutrinoless double beta decay. Current experimental limits on this process impose severe constraints on the inverted ordering realization of the $3+1$ scenario, excluding the small breaking regime within the parameter ranges considered. In contrast, the normal ordering case remains viable and predicts values of the effective Majorana mass that lie within the sensitivity reach of upcoming neutrinoless double beta decay experiments.
  
\end{abstract}

\section{Introduction}

The current paradigm of neutrino physics is built upon the PMNS (Pontecorvo-Maki-Nakagawa-Sakata) matrix, which parameterizes the mixing between flavor eigenstates and mass eigenstates. The squared-mass differences, $\Delta m^2_{21} \approx 7.5 \times 10^{-5} \, \text{eV}^2$ and $|\Delta m^2_{31}| \approx 2.5 \times 10^{-3} \, \text{eV}^2$, dictate the oscillation frequencies observed in solar, atmospheric, reactor, and accelerator experiments which have reached unprecedented precision \cite{ParticleDataGroup:2024cfk,Esteban:2024eli,Capozzi:2025wyn,deSalas:2020pgw}. Despite the robustness of the $3\nu$ framework, it remains an incomplete description as it does not account for the absolute mass scale, the ordering (normal or inverted), and the potential violation of CP symmetry in the leptonic sector.

While the three-neutrino framework remains the baseline description of flavor oscillations, firmly established by experimental data, the possibility of additional sterile neutrino states has been motivated by a series of anomalies observed in short-baseline (SBL) experiments, most notably the appearance signals reported by
LSND and MiniBooNE~\cite{LSND:2001aii,MiniBooNE:2013uba,MiniBooNE:2020pnu}. Independent evidence has arisen from disappearance anomalies in gallium neutrino-source experiments~\cite{GALLEX:1997lja,SAGE:1998fvr,Barinov:2021asz}, while the reevaluation of reactor antineutrino fluxes provided an additional indication of SBL $\nu_e$ and $\bar{\nu}_e$ disappearance, giving rise to the so-called reactor antineutrino anomaly (RAA)~\cite{Mention:2011rk,Giunti:2021kab,Giunti:2023kyo,Elliott:2023cvh,Giunti:2022btk}. These anomalies have motivated the consideration of an additional oscillation scale characterized by a mass-squared difference of order $\Delta m^2_{41}\sim 1\,\mathrm{eV}^2$, providing a phenomenological motivation for extending the neutrino sector by a fourth, predominantly sterile, mass eigenstate within the minimal $3+1$ framework~\cite{Gariazzo:2019cym,Danilov:2022str,Acero:2022wqg}.

Global analyses of the $3+1$ framework have found that the regions preferred by appearance data are in significant tension with those allowed by disappearance data~\cite{Gariazzo:2017fdh}. A pragmatic approach that excludes, for instance, the low-energy MiniBooNE excess from the global combination has been implemented in order to alleviate this tension~\cite{Gariazzo:2017fdh}. Although a recent MicroBooNE analysis excluded, at the $95\%$ confidence level, the interpretation of the LSND and MiniBooNE anomalies in terms of a single light sterile neutrino, while also excluding a significant fraction of the parameter space relevant to the gallium anomaly~\cite{MicroBooNE:2025nll}, modern simulation-based inference (SBI) evaluations indicate a reduced but non-negligible disagreement between the MiniBooNE and MicroBooNE data sets~\cite{Woodward:2026jnx}. In addition, a recent reevaluation based on the most complete reactor antineutrino flux model has indicated a revival of the RAA, whose simplest explanation remains active-sterile neutrino oscillations~\cite{Giunti:2026uaf}. Hence, the tension among the different data sets relevant to the $3+1$ sterile neutrino framework remains persistent, and the scenario is thus narrowed rather than closed.

A sterile neutrino ($\nu_s$), a singlet under the Standard Model gauge group $SU(2)_L\times U(1)_Y$, arises naturally in extensions of the lepton sector, and in particular in the seesaw mechanisms invoked to explain the smallness of the active neutrino masses\footnote{It should be stressed, however, that an eV-scale singlet is not a generic prediction of the canonical type-I seesaw, but requires model constructions with an approximate lepton-number symmetry~\cite{Barry:2011wb,Zhang:2011vh}.}. The $3+1$ scenario adds a single mass eigenstate $\nu_4$, with $m_4$ heavier than $m_{1,2,3}$, so that flavor and mass states are related by a $4\times4$ unitary matrix; the new splitting $\Delta m^2_{41}\sim 1\,\mathrm{eV}^2$ accounts for high-frequency oscillations at small $L/E$ while leaving the established three-flavor phenomenology essentially untouched. From a phenomenological standpoint, the constraints arising from forthcoming results may nonetheless prove valuable in discriminating among neutrino mass models, in particular those predicting specific values of the mixing angles and those bearing on $CP$-violation observables within the $3+1$ framework~\cite{Klop:2014ima,Rivera-Agudelo:2015vza,Becerra-Garcia:2021gcd}. Mixing patterns such as cobimaximal mixing are linked to discrete symmetries relating specific neutrino flavors, namely the $\mu-\tau$ reflection symmetry. In the three-neutrino framework, the values predicted in the exact symmetry limit differ from those determined experimentally, so that the symmetry may be realized only in an approximate way. Previous works have explored the breaking of similar flavor patterns, such as the $\mu-\tau$ interchange symmetry, induced by an additional sterile neutrino state, and have extracted information about the model parameters from the existing experimental bounds~\cite{Rivera-Agudelo:2015vza,Becerra-Garcia:2021gcd}. Early attempts to explore the $\mu-\tau$ interchange symmetry were carried out by imposing the exact symmetry on specific columns of the mixing matrix~\cite{Chakraborty:2019rjc}.

In this work we adopt a phenomenological approach to constrain the Dirac and Majorana $CP$-violating phases of the $3+1$ neutrino scenario by linking the additional state to the breaking of the $\mu-\tau$ reflection symmetry of the neutrino mass matrix. The paper is organized as follows. In Sec.~\ref{mutau} we review the essentials of cobimaximal mixing and its relation to the reflection symmetry. In Sec.~\ref{sterile} we connect the $3+1$ scenario to the breaking of this symmetry and derive analytical expressions for the relevant experimental observables. In Sec.~\ref{results} we present the results of our analysis, together with their implications for the neutrinoless double beta decay amplitude. Finally, in Sec.~\ref{sec:conclusions} we summarize our findings and offer concluding remarks.

\section{$\mu-\tau$ reflection symmetry}\label{mutau}

An especially important consequence of $\mu-\tau$ reflection symmetry is its connection with the cobimaximal mixing pattern~\cite{Ma:2015pma,Grimus:2003yn}, which is characterized by a
nonvanishing reactor angle $\theta_{13}$, maximal atmospheric mixing $\theta_{23}=\pi/4$, and maximal $CP$ violation with $\delta_{CP}=\pm\pi/2$. Unlike the earlier tribimaximal mixing
paradigm \cite{Harrison:2002er,Grimus:2012hu}, which predicted a vanishing reactor angle, cobimaximal mixing remains fully compatible with present oscillation data~\cite{Esteban:2024eli}. Consequently, it has become one of the most attractive benchmark scenarios for constructing realistic models of lepton flavor. In many frameworks, cobimaximal mixing arises naturally as a direct manifestation of $\mu-\tau$ reflection symmetry, which makes the latter a fundamental organizing principle for understanding the structure of the PMNS matrix~\cite{Ma:2015pma,Xing:2022uax}. Adopting a phase convention in which the $\mu$ and $\tau$ rows are rephased to render the reflection property manifest\footnote{This choice is unobservable in oscillation experiments, since it can be absorbed into the charged lepton fields.}, the mixing matrix takes the cobimaximal form
\begin{equation}\label{eq:cbm3x3}
U_{\mathrm{CBM}}=
\begin{pmatrix}
c_{12}c_{13} &
s_{12}c_{13} &
\pm i\,s_{13}
\\
-\dfrac{1}{\sqrt{2}}\left(s_{12} \pm i\,c_{12}s_{13}\right) &
\dfrac{1}{\sqrt{2}}\left(c_{12} \mp i\,s_{12}s_{13}\right) &
-\dfrac{c_{13}}{\sqrt{2}}
\\
-\dfrac{1}{\sqrt{2}}\left(s_{12} \mp i\,c_{12}s_{13}\right) &
\dfrac{1}{\sqrt{2}}\left(c_{12} \pm i\,s_{12}s_{13}\right) &
\dfrac{c_{13}}{\sqrt{2}}
\end{pmatrix},
\end{equation}
where $c_{ij}\equiv\cos\varphi_{ij}$ and $s_{ij}\equiv\sin\varphi_{ij}$, and the upper(lower) sign correspond with the selection $\delta_{CP}=-\pi/2(+\pi/2)$, respectively, and vanishing Majorana phases. Throughout this work we reserve the symbols $\varphi_{ij}$ and $\delta_{CP}$ for the parameters of the symmetric ($3\times3$) limit, and denote by $\theta_{ij}$ and $\delta_{13}$ the effective mixing angles and phase extracted from global fits under the three neutrino hypothesis~\cite{Esteban:2024eli}.   

The mixing matrix presented in Eq. (\ref{eq:cbm3x3}) gives rise to the well-known $\mu-\tau$ reflection symmetry of the neutrino mass matrix, associated with the partially flavor-changing
$CP$ transformations $\nu_{eL}\rightarrow(\nu_{eL})^{c}$, $\nu_{\mu L}\rightarrow(\nu_{\tau L})^{c}$ and $\nu_{\tau L}\rightarrow(\nu_{\mu L})^{c}$, where $(\nu_{\alpha L})^{c}$ denotes the charge conjugate of the corresponding state. Indeed, requiring invariance of the Majorana mass term under these transformations constrains the mass matrix, written in the flavor basis as $M_{0}=U_{\mathrm{CBM}}\,\mathrm{diag}(m_{1},m_{2},m_{3})\,U_{\mathrm{CBM}}^{T}$, to exhibit a reflection property between the $\mu$ and $\tau$ labels,
\begin{equation}
\label{eq:massmatrix}
M_{0} = \left(
\begin{array}{c@{\quad}c@{\quad}c}
m_{ee}        & m_{e\mu}    & m_{e\mu}^{*}
\\
m_{e\mu}      & m_{\mu\mu}  & m_{\mu\tau}
\\
m_{e\mu}^{*}  & m_{\mu\tau} & m_{\mu\mu}^{*}
\end{array}
\right),
\end{equation}
which is summarized by the four \textit{reflection} conditions \cite{Xing:2022uax}
\begin{equation}\label{eq:relations}
m_{ee}=m_{ee}^{*},\qquad
m_{\mu\tau}=m_{\mu\tau}^{*},\qquad
m_{e\mu}=m_{e\tau}^{*},\qquad \text{and}\qquad
m_{\mu\mu}=m_{\tau\tau}^{*}.
\end{equation}
The first two conditions state that the $ee$ and $\mu\tau$ entries are real, whereas the remaining two relate pairs of entries by complex conjugation. This symmetry, therefore, removes four of the phases that would otherwise be free in a general complex symmetric matrix, and it is this reduction that underlies the predictive power of the scenario.

Despite its success, current global fits indicate that the measured values of the mixing parameters exhibit small but potentially significant deviations from the exact predictions of $\mu-\tau$ reflection symmetry, most notably a preference for $\theta_{23}$ away from $\pi/4$ and for $\delta_{CP}$ away from $-\pi/2$~\cite{Esteban:2024eli}. This observation motivates the study of symmetry breaking effects and their phenomenological consequences. Various mechanisms have been proposed to generate controlled departures from the exact symmetric limit, including explicit breaking at the level of the mass matrix, radiative corrections induced by the renormalization group running between the flavor symmetry scale and the electroweak scale, higher dimensional operators, and residual breaking effects originating in the flavor sector~\cite{Xing:2022uax}.Understanding these deviations is particularly relevant in the era of precision neutrino physics. Next-generation long-baseline experiments such as DUNE~\cite{DUNE:2020jqi} and Hyper-Kamiokande \cite{Hyper-KamiokandeProto-:2015xww} will substantially improve the determination of $\theta_{23}$ and $\delta_{\rm CP}$, while JUNO~\cite{JUNO:2022mxj} will provide complementary precision measurements of $\theta_{12}$ and the neutrino mass-squared differences. Together, these measurements will enable stringent tests of flavor models based on specific symmetry scenarios.

Such deviations may also be linked to the corresponding $\mu-\tau$ reflection symmetry of the mass matrix. Scenarios of this kind have been explored within the three neutrino framework, yielding hints on the $CP$-violating phases and on the effective mass governing neutrinoless double beta decay. Beyond three flavors, a few works addressing the $3+1$ scheme have examined the breaking of the $\mu-\tau$ interchange symmetry~\cite{Rivera-Agudelo:2015vza,Becerra-Garcia:2021gcd} and of the reflection symmetry within different approaches~\cite{Chakraborty:2019rjc}. In contrast to the $\mu-\tau$ interchange symmetry, which is not compatible with a nonvanishing $\theta_{13}$ in its exact limit, the reflection symmetry accommodates the measured value of the reactor angle while still constraining the phases, which makes it the more suitable starting point for the analysis presented below.

It is worth delineating our approach with respect to earlier analysis regarding the $3+1$ scenario in the context of the $\mu-\tau$ reflection symmetry. In Ref. \cite{Chakraborty:2019rjc}, the reflection condition is directly imposed on the $4\times4$ mixing matrix, either globally or column by column, which results in correlations among the mixing parameters depending on the particular choice. In this work, by contrast, we will start from the mass matrix, where the coupling to the fourth state will be treated as the source of symmetry violation. The sterile neutrino would therefore be the sole agent of symmetry deviations, so that the breaking is genuine rather than merely apparent. Consequently, the four reflection relations no longer hold in the active sector, and the physical mixing angles and phases would depart from their symmetric values.


\section{3+1 neutrino framework}\label{sterile}

Within the Standard Model, the only available source of breaking of the $\mu-\tau$ reflection symmetry is the mass splitting between the muon and the tau lepton. The associated radiative corrections, however, have been shown to be too small to account for the observed departures from the symmetric limit \cite{Gomez-Izquierdo:2007scz}, which suggests that the breaking mechanism must be sought beyond the Standard
Model. Among the candidates, a sterile neutrino has been examined in the context of both the $\mu-\tau$ interchange and the $\mu-\tau$ reflection symmetries \cite{Rivera-Agudelo:2015vza,Becerra-Garcia:2021gcd,Chakraborty:2019rjc}. As noted above, the experimentally allowed ranges deviate from the exact predictions while remaining compatible with them at the three standard deviation level,
so that the symmetry, if present, can only be slightly broken. If the fourth state is heavier than the active ones by more than an order of magnitude in mass, its effects on the low-energy mass matrix are suppressed by the ratio $m_{\nu}/m_{s}$, and the induced breaking is naturally small. The magnitude of the deviations is thus not an independent input but is controlled by the active-sterile mixing.

The most general $4\times4$ Majorana mass matrix with a $\mu-\tau$ symmetry in the active sector sector is expressed as
\begin{equation}
\label{eq:mass4}
M_{\nu} = \left( \begin{array}{c@{\quad}c@{\quad}c}
M_{0}  & \bm{\alpha} m_s 
\\
\bm{\alpha}^T m_s & m_s
\end{array}  \right),
\end{equation} 
where $M_{0}$ stands as the $3\times 3$ symmetric matrix in Eq. (\ref{eq:massmatrix}), $m_s$ is the sterile mass, and $\bm{\alpha}^T = (\alpha_e,\alpha_\mu,\alpha_\tau)$ represents the dimensionless sterile to active neutrino mixing parameter. In order to not spoil the well-known $3\nu$ scheme, the mixing between active and sterile neutrinos must be small. In addition, for the sake of simplicity, let us assume these parameters to be real\footnote{It is also possible to consider complex mixing parameters \cite{Becerra-Garcia:2021gcd}, but, as we will show later, this would increase the number of phases in the mass matrix and hence the number of breaking parameters of our analysis.}.

After decoupling $\nu_s$, the small-scale seesaw framework allows us to express, at low energies, the effective active neutrino mass terms as \cite{Barry:2011wb}
\begin{equation}\label{eq:Mdecoup}
m'_{\ell n} \simeq m_{\ell n} - \alpha_\ell \alpha_{n} m_s ~,
\end{equation}
where $\ell,n = e,\mu,\tau$. It is important to note from Eq.~(\ref{eq:Mdecoup}) that the active neutrino mass matrix will present deviations from the reflection symmetry induced by the
$\alpha_{\ell}$ parameters, which reflects that the associated mixings would also be modified. In this framework we have further assumed $m_4 \sim m_s \sim 1~{\rm eV}$, so that the sterile scale exceeds the active entries of $M_0$ by roughly two orders of magnitude. The induced breaking is therefore amplified with respect to the naive size of the couplings, a point that we quantify below.

The former two conditions in Eq.~(\ref{eq:relations}) are fulfilled by assuming the $\alpha_{\ell}$ to be real. Hence, the breaking of the remaining conditions can be parameterized by defining two dimensionless parameters as in \cite{Rivera-Agudelo:2024vdn}, $\hat \delta = (m'_{e\tau} -
m'^*_{e\mu})/m'^*_{e\mu}$ and $\hat \epsilon = (m'_{\tau \tau} - m'^*_{\mu \mu})/m'^*_{\mu \mu}$, which allows us to express
\begin{eqnarray}\label{eq:parameters}
\hat \delta &\simeq& \frac{m_s\alpha_e(\alpha_\mu - \alpha_{\tau})}
{m^*_{e\mu}-m_s\alpha_e \alpha_\mu}~, \nonumber \\
\hat \epsilon &\simeq& \frac{m_s(\alpha^2_\mu-\alpha^2_\tau)}
{m^*_{\mu\mu}-m_s\alpha^2_{\mu}}.
\end{eqnarray}
It is worth noticing that different parameter combinations may lead to different deviations from the $\mu-\tau$ symmetry. Indeed, a closer inspection of Eq.~(\ref{eq:parameters}) reveals that the breaking is not governed by the overall magnitude of the couplings, but by their difference in the $\mu-\tau$ sector. We observe that both numerators are proportional to $\alpha_\mu - \alpha_\tau$, since $\alpha^2_\mu-\alpha^2_\tau = (\alpha_\mu+\alpha_\tau)(\alpha_\mu-
\alpha_\tau)$. Consequently, the symmetric limit $\hat\delta = \hat\epsilon = 0$ is recovered whenever $\alpha_\mu = \alpha_\tau$, regardless of how large the individual couplings
are, whereas a hierarchical choice such as $\alpha_\tau \ll \alpha_\mu$ induces an $\mathcal{O}(1)$ breaking even for small couplings. This is a welcome feature since the condition
$\alpha_\mu = \alpha_\tau$ is precisely the statement that the vector $\bm{\alpha}$ is itself invariant under the $\mu-\tau$ exchange, so that the reflection symmetry of the underlying $4\times4$ mass matrix is exact and the sterile state does not spoil it. The parameters $\hat\delta$ and $\hat\epsilon$ thus measure the departure of $\bm{\alpha}$ from this
extended symmetric limit, and it is natural to organize the analysis in terms of the combinations
\begin{equation}\label{eq:combinations}
\bar\alpha \equiv \frac{\alpha_\mu+\alpha_\tau}{2}~, \qquad
\Delta\alpha \equiv \alpha_\mu-\alpha_\tau~,
\end{equation}
with $\Delta\alpha/\bar\alpha \ll 1$ as the relevant smallness condition. In this notation, Eq.~(\ref{eq:parameters}) becomes, to leading order in the active-sterile couplings,
\begin{equation}\label{eq:parameters2}
\hat\delta \simeq \frac{m_s\alpha_e\,\Delta\alpha}{m^*_{e\mu}}~,
\qquad
\hat\epsilon \simeq \frac{2m_s\bar\alpha\,\Delta\alpha}{m^*_{\mu\mu}}~,
\end{equation}
which exhibits the announced amplification. Given that $|m^*_{e\mu}|\sim\sqrt{\Delta m^2_{21}}$ and
$|m^*_{\mu\mu}|\sim\sqrt{\Delta m^2_{31}}$ are of order $10^{-2}~{\rm eV}$ while $m_s\sim1~{\rm eV}$, the prefactors $m_s/|m^*_{\rho\sigma}|$ are of order $10^{2}$, and a breaking at the ten percent level already requires $\Delta\alpha \lesssim 10^{-2}$ for $\alpha_e \sim \bar\alpha \sim 0.1$. We emphasize that this constraint applies only to the difference, allowing couplings to reach the values $|U_{\alpha4}|^2\sim10^{-2}$ suggested by the short-baseline anomalies without compromising the approximate symmetry, which is what makes the present scenario phenomenologically viable. The ratio $\hat\epsilon/\hat\delta \simeq 2\bar\alpha m^*_{e\mu}/(\alpha_e m^*_{\mu\mu})$ is moreover independent of $\Delta\alpha$, so that the relative size of the two breaking parameters is fixed by the flavor structure of $\bm{\alpha}$ and not by the overall amount of breaking. This
correlation, which we exploit in Sec.~\ref{results}, is a distinctive prediction of treating the sterile state as the sole breaking agent.

To investigate the deviations in the mixings it is mandatory to obtain the $4\times4$ mixing matrix that matches the observed bounds on active neutrino masses. We should note that Eq.~(\ref{eq:mass4}) can be diagonalized in a perturbative approximation by splitting the mass matrix in the form $M_\nu = {\cal M}_0 + {\cal V}$, where
\begin{equation}
{\cal M}_0 =
\begin{pmatrix}
M_0 & 0 \\
0 & m_s
\end{pmatrix} ~~ {\rm and } ~~
{\cal V} =
\begin{pmatrix}
0 & \bm{\alpha} m_s \\
\bm{\alpha}^T m_s & 0
\end{pmatrix}~.
\end{equation}
Here, $M_0$ stands for the $3\times3$ mass matrix in Eq.~(\ref{eq:massmatrix}), and $m_s$ remains the sterile mass. The unperturbed $\mu-\tau$ symmetric matrix $M_0$ can be directly diagonalized by
the cobimaximal matrix in Eq.~(\ref{eq:cbm3x3}). Note that this splitting is legitimate as a perturbative expansion because the off-diagonal blocks are suppressed relative to the sterile eigenvalue by the couplings themselves, $\|{\cal V}\|/\|{\cal M}_0\| \sim |\bm{\alpha}|$, and the two sectors are widely separated in mass, $m_s \gg m_{1,2,3}$. The expansion parameter is
therefore $|\alpha_\ell|$, and the resulting series is controlled independently of whether the breaking parameters $\hat\delta,\hat\epsilon$ are small. Hence, in what follows we will consider that the eigenvalues of $M_0$ give, at lowest order, the active mass eigenvalues $(m_1,m_2,m_3)$, which, as we will show later, can be expressed in terms of the solar and atmospheric scales, whereas $m_4 \approx m_s$.

A straightforward calculation allows us to obtain the eigenvectors that diagonalize $M_\nu$ up to second order in perturbation theory, as suggested by the corrections to the active mass matrix entries in Eq.~(\ref{eq:Mdecoup}), and thereby to express the PMNS matrix entries,
organized by columns, in the following way:
 
\begin{align}\label{eq:column1}
U_{e1} &= c_{12} c_{13} + \frac{m_{14} X}{4} \left( i m_{13} Y s_{13} - m_{12} Z c_{13} s_{12} \right) \nonumber \\
U_{\mu 1} &= \frac{1}{\sqrt{2}} \left[ -(s_{12} + i c_{12} s_{13}) + \frac{m_{14} X}{4} \left( -m_{13} Y c_{13} + m_{12} Z (c_{12} - i s_{12} s_{13}) \right) \right] \nonumber \\
U_{\tau 1} &= \frac{1}{\sqrt{2}} \left[ -s_{12} + i c_{12} s_{13} + \frac{m_{14} X}{4} \left( m_{13} Y c_{13} - m_{12} Z (c_{12} + i s_{12} s_{13}) \right) \right] \nonumber \\
U_{s1} &= -\frac{1}{2} m_{14} X 
\end{align}

\begin{align}\label{eq:column2}
U_{e2} &= c_{13} s_{12} + \frac{m_{24} Z^*}{4} \left( m_{12} c_{12} c_{13} X^* - i m_{23} s_{13} Y \right) \nonumber \\
U_{\mu 2} &= \frac{1}{\sqrt{2}} \left[ (c_{12} - i s_{12} s_{13}) + \frac{m_{24} Z^*}{4} \left( m_{23} c_{13} Y - m_{12} X^* (s_{12} + i c_{12} s_{13}) \right) \right] \nonumber \\
U_{\tau 2} &= \frac{1}{\sqrt{2}} \left[ (c_{12} + i s_{12} s_{13}) - \frac{m_{24} Z^*}{4} \left( m_{23} c_{13} Y + m_{12} X^* (s_{12} - i c_{12} s_{13}) \right) \right] \nonumber \\
U_{s 2} &= \frac{1}{2}  m_{24} Z^*
\end{align}

\begin{align}\label{eq:column3}
U_{e3} &= i s_{13} + \frac{m_{34} c_{13} Y^*}{4} \left( -m_{13} c_{12} X^* + m_{23} s_{12} Z \right) \nonumber \\
U_{\mu 3} &= \frac{1}{\sqrt{2}} \left[ -c_{13} + \frac{m_{34} Y^*}{4} \left( m_{13} X^* (s_{12} + i c_{12} s_{13}) + m_{23} Z (c_{12} - i s_{12} s_{13}) \right) \right] \nonumber \\
U_{\tau 3} &= \frac{1}{\sqrt{2}} \left[ c_{13} + \frac{m_{34} Y^*}{4} \left( m_{13} X^* (s_{12} - i c_{12} s_{13}) + m_{23} Z (c_{12} + i s_{12} s_{13}) \right) \right] \nonumber \\
U_{s 3} &= -\frac{1}{2} m_{34} Y^* 
\end{align}

\begin{align}\label{eq:column4}
U_{e4} &= \frac{1}{2} \left( i s_{13} Y m_{34} + c_{12} c_{13} X^* m_{14} - s_{12} c_{13} Z m_{24} \right) \nonumber \\
U_{\mu 4} &= -\frac{1}{2\sqrt{2}} \left[ c_{13} Y m_{34} + (s_{12} + i c_{12} s_{13}) X^* m_{14} + (c_{12} - i s_{12} s_{13}) Z m_{24} \right] \nonumber \\
U_{\tau 4} &= \frac{1}{2\sqrt{2}} \left[ c_{13} Y m_{34} - (s_{12} - i c_{12} s_{13}) X^* m_{14} - (c_{12} + i s_{12} s_{13}) Z m_{24} \right] \nonumber \\
U_{s 4} &= 1
\end{align}
where $m_{ij} = \frac{m_4}{m_i - m_j}$, with $i,j=1,2,3,4$. We have assumed the following definitions: 
\begin{align}
X &= 2\sqrt{2}\bar{\alpha} s_{12} + c_{12} \left[ -2\alpha_{e} c_{13} + i\sqrt{2}\Delta\alpha s_{13} \right] \nonumber \\
Y &= \sqrt{2}\Delta\alpha c_{13} + 2i\alpha_{e} s_{13} \\
Z &= 2\sqrt{2}\bar{\alpha} c_{12} + s_{12} \left[ 2\alpha_{e} c_{13} + i\sqrt{2}\Delta\alpha s_{13} \right] ~. \nonumber
\end{align}
It can be seen, in this notation, that the PMNS matrix elements of the active sector show as a first term the underlying combination of mixing angles characteristic of the cobimaximal pattern, and that products of the $X$, $Y$ and $Z$ factors induce corrections of ${\cal O}(\alpha_{\ell}^2)$, as was expected. The deviations from the mixing pattern may be small depending on the size of the active-sterile couplings, which can be directly estimated from the bounds on the elements in Eq.~(\ref{eq:column4}). These bounds, as we can see from Eq.~(\ref{eq:Mdecoup}), should also change the symmetry relations of the mass matrix. On the other hand, the active-sterile elements
remain of ${\cal O}(\alpha_{\ell})$, while the pure sterile element is of ${\cal O}(1)$. Finally, it is straightforward to verify that, in the limit of null coupling parameters, the $X$, $Y$ and $Z$ factors vanish, leading to a cobimaximal mixing matrix in the active sector with no active-sterile mixing.

The matrix elements in Eqs.~(\ref{eq:column1})--(\ref{eq:column4}) may be compared with the general parametrization of the $3+1$ scheme \cite{Barry:2011wb}, obtained from
\begin{equation}\label{eq:pmns4x4}
U = R(\theta_{34})\,R(\theta_{24},\delta_{24})\,R(\theta_{14},\delta_{14})\,
R(\theta_{23})\,R(\theta_{13},\delta_{13})\,R(\theta_{12})\,P,
\end{equation}
where $R(\theta_{ij})$ describes a rotation in the $ij$ plane and $\delta_{ij}$ a Dirac $CP$-violating phase located in the corresponding rotation, in such a way that $\delta_{13}$ can be linked to the $CP$-violating phase of the standard PMNS matrix of the three active neutrinos~\cite{Dev:2019pdg}. Throughout this section the complex rotations are defined with the phase factor $e^{-i\delta_{ij}}$ in the $(i,j)$ entry; we stress this convention because the sign of the rephasing invariants introduced below depends on it. Here, $P={\rm diag}(1, e^{i\alpha/2},e^{i(\beta/2+\delta_{13})}, e^{i(\gamma/2+\delta_{14})})$ is a diagonal matrix written in terms of the Majorana phases relevant for neutrinoless double beta decay. We emphasize that the angles $\theta_{ij}$ and phases $\delta_{ij}$ entering Eq.~(\ref{eq:pmns4x4}) are the physical, experimentally accessible parameters, and are to be distinguished from the symmetric-limit quantities $\varphi_{ij}$ and $\delta_{CP}$ introduced in Sec.~\ref{mutau}.

Finally, the relevant mixing angles in the general parametrization can be obtained from the PMNS matrix elements in Eqs.~(\ref{eq:column1})--(\ref{eq:column4}) through the following relations
\begin{equation}
\label{eq:mixings}
\begin{aligned}
\sin^{2}\theta_{14} &= |U_{e4}|^{2},\\
\sin^{2}\theta_{24} &= \frac{|U_{\mu4}|^{2}}{1-|U_{e4}|^{2}},\\
\sin^{2}\theta_{34} &= \frac{|U_{\tau4}|^{2}}
{1-|U_{e4}|^{2}-|U_{\mu4}|^{2}},\\
\sin^{2}\theta_{13} &= \frac{|U_{e3}|^{2}}{1-|U_{e4}|^{2}},\\
\sin^{2}\theta_{12} &= \frac{|U_{e2}|^{2}}
{1-|U_{e4}|^{2}-|U_{e3}|^{2}},\\
\sin^{2}\theta_{23} &=
\frac{|U_{\mu3}|^{2}\left(1-|U_{e4}|^{2}\right)
-|U_{e4}|^{2}|U_{\mu4}|^{2}}
{1-|U_{e4}|^{2}-|U_{\mu4}|^{2}}
\\
&\quad+
\frac{|U_{e1}U_{\mu1}+U_{e2}U_{\mu2}|^{2}
\left(1-|U_{e4}|^{2}\right)}
{\left(1-|U_{e4}|^{2}-|U_{e3}|^{2}\right)
\left(1-|U_{e4}|^{2}-|U_{\mu4}|^{2}\right)}.
\end{aligned}
\end{equation}
These relations hold exactly, that is, to all orders in the active-sterile couplings, and reduce to the familiar three-neutrino expressions in the limit $|U_{\alpha4}|\rightarrow0$. Note that the nonunitarity of the active block manifests itself in the denominators, which normalize each modulus to the subspace orthogonal to the states already extracted; this is the reason why $\theta_{12}$ and $\theta_{23}$ acquire increasingly involved expressions, and why the naive identification $\sin^2\theta_{ij}=|U_{\alpha i}|^2$ would be inconsistent beyond leading order.

Predictions for the mixing angles can be extracted from specific matrix elements and directly confronted with current global analyses of neutrino oscillation data. Within our model, we identify nine free parameters: three active-sterile couplings $\alpha_\ell$, two mixing angles $\varphi_{12}$ and $\varphi_{13}$, and four mass eigenvalues (see Eqs.~(\ref{eq:column1})--(\ref{eq:column4})). In what follows, we perform a scan of the model parameter space by imposing the experimental constraints on the active and active-sterile mixing parameters~\cite{Esteban:2024eli,Gariazzo:2017fdh}. In this way, the nine model parameters can be confronted with nine experimental inputs, namely the three active mixing angles, three bounds on the active-sterile matrix elements, and three mass-squared differences. It should be stressed, however, that this counting is not a one-to-one matching sector by sector: the two symmetric-limit angles are confronted with three measured active angles, so that $\theta_{23}$ is genuinely predicted rather than fitted, whereas the
four mass eigenvalues are constrained by only three mass-squared splittings, leaving the lightest mass as a free parameter which we vary within the range allowed by cosmological and kinematic bounds. Within this framework, in addition to the mixing angles defined in Eq.~(\ref{eq:mixings}), the six $CP$-violating phases (three Dirac and three Majorana) appearing in Eq.~(\ref{eq:pmns4x4}) can also be predicted, since none of them is used as an input of the scan.

In the $3+1$ neutrino-mixing scheme, the Dirac $CP$-violating phases can be extracted from the rephasing-invariant quartets of mixing-matrix elements,
whose imaginary parts constitute the generalization of the Jarlskog invariant to a four-neutrino framework and, once divided by the corresponding products the mixing angles, yield 
\begin{align}\label{eq:diracs}
\sin\delta_{13}
&=
\frac{
    \operatorname{Im}\!\left(
        U_{e1}U_{\mu3}U_{e3}^{*}U_{\mu1}^{*}
    \right)
}{
    c_{12}s_{12}c_{13}^{2}s_{13}
    c_{23}s_{23}c_{14}c_{24}
},
\\[6pt]
\sin\delta_{14}
&=
\frac{
    \operatorname{Im}\!\left(
        U_{e1}U_{\mu4}U_{e4}^{*}U_{\mu1}^{*}
    \right)
}{
    c_{12}s_{12}c_{13}c_{14}^{2}s_{14}
    c_{24}s_{24}
},
\\[6pt]
\sin\delta_{24}
&=
\frac{
    \operatorname{Im}\!\left(
        U_{\mu2}U_{\tau4}U_{\mu4}^{*}U_{\tau2}^{*}
    \right)
}{
    c_{12}c_{23}s_{23}c_{13}
    c_{24}^{2}s_{24}c_{34}s_{34}
}.
\end{align}
These quantities have the generic form and constitute the generalization of the Jarlskog invariant to a four-neutrino framework. These Dirac phases can contribute to neutrino oscillation probabilities and generate $CP$-violating differences between neutrino and antineutrino oscillations. The precise expressions depend on the ordering of the rotations and on the phase convention adopted for the mixing matrix ~\cite{Schechter:1980gr,Kopp:2013vaa}.

On the other hand, the three Majorana phases can be extracted from rephasing-invariant combinations of the elements in the electron row of the mixing matrix. Taking $U_{e1}$ as the phase reference, they are defined as
\begin{align}\label{eq:majoranas}
\sin\alpha &=
\frac{{\rm Im}\left[U_{e2}^{\,2}\left(U_{e1}^{*}\right)^{2}\right]}
{|U_{e1}|^{2}|U_{e2}|^{2}},
\\
\sin\beta &=
\frac{{\rm Im}\left[U_{e3}^{\,2}\left(U_{e1}^{*}\right)^{2}\right]}
{|U_{e1}|^{2}|U_{e3}|^{2}},
\\
\sin\gamma &=
\frac{{\rm Im}\left[U_{e4}^{\,2}\left(U_{e1}^{*}\right)^{2}\right]}
{|U_{e1}|^{2}|U_{e4}|^{2}}.
\end{align}
The squared matrix elements appear because the Majorana phases enter lepton number violating observables, such as neutrinoless double beta decay, through combinations of the form $U_{ei}^{2}m_i$. These phase differences are invariant under a common rephasing of the electron charged-lepton
field\footnote{For discussions of Majorana phases and lepton number violating observables in sterile neutrino scenarios, see, for example, Ref.~\cite{Barry:2011wb}.}.

It is worth noting that, by controlling the deviations from the symmetric mixing pattern through the relations given in Eq.~(\ref{eq:parameters}), the model parameter space can be further reduced and, consequently, so can the full set of physical parameters. Since both $\hat\delta$ and $\hat\epsilon$ are proportional to $\Delta\alpha$, requiring the breaking to remain below a given level restricts the scan to a neighborhood of the surface $\alpha_\mu=\alpha_\tau$ in coupling space, which in turn could confine the predicted values of the model and mixing parameters to well-defined regions. This correlation between the amount of symmetry breaking and the location of the physical parameters is the central result of our analysis, and is examined numerically in Sec.~\ref{results}.

\section{Results}\label{results}

In order to evaluate the expressions in Eqs.~(\ref{eq:parameters}) and~(\ref{eq:column1})-(\ref{eq:column4}), which account for the deviations from reflection symmetry in the neutrino mass matrix and from the cobimaximal pattern, respectively, a specific neutrino mass ordering must be assumed. The four mass eigenvalues can be parametrized in terms of the measured mass-squared differences and the lightest neutrino mass ($m_0$), where the sterile state is taken to be above the active states. For normal ordering (NO), the three active states satisfy $m_1<m_2<m_3$, with $m_1=m_0$, such that $m_2=\sqrt{m_0^2+\Delta m_{\text{SOL}}^2}$, and $m_3=\sqrt{m_0^2+\Delta m_{\text{ATM}}^2}$, using the solar and atmospheric mass-squared splittings, respectively. The sterile state is then usually parametrized through the additional splitting $\Delta m_{41}^2$, namely $m_4=\sqrt{m_0^2+\Delta m_{41}^2}$. For inverted ordering (IO), we have $m_3<m_1<m_2$, and the lightest state is instead $m_3=m_0$. In this case, one may write $m_1=\sqrt{m_0^2-\Delta m_{\text{ATM}}^2}$ and $m_2=\sqrt{m_0^2-\Delta m_{\text{ATM}}^2+\Delta m_{\text{SOL}}^2}$, while the sterile mass is given by $m_4=\sqrt{m_0^2-\Delta m_{\text{ATM}}^2+\Delta m_{41}^2}$. 

In the following analysis, we adopt the results reported in Ref.~\cite{Esteban:2024eli} for the neutrino mixing angles and those of Ref.~\cite{Gariazzo:2017fdh} for the sterile sector, considering, in both cases, the corresponding $3\sigma$ allowed intervals. The mass-squared differences are fixed at their central values since previous studies have shown that their variation produces only minor effects on the final estimates \cite{Rivera-Agudelo:2024vdn}. The numerical input values for the active mixing used in our analysis are summarized in Table~\ref{tab:data}.

\begin{table}
\centering
\caption{Values of the mixing angles ($3\sigma$ interval) and squared mass differences ($1\sigma$) used in the numerical evaluations \cite{Esteban:2024eli}.}
\label{tab:data}
\begin{tabular}{lcc}
\toprule
Parameter & Normal Ordering (NO) & Inverted Ordering (IO) \\
\midrule
$\sin^2 \theta_{12}$  & $0.2893 \to 0.3295$ & $0.2893 \to 0.3295$ \\
$\sin^2 \theta_{23}$  & $0.435 \to 0.584$ & $0.439 \to 0.584$ \\
$\sin^2 \theta_{13}$  & $0.02064 \to 0.02418$ & $0.02093 \to 0.02441$ \\
\addlinespace
$\frac{\Delta m^2_{\text{SOL}}}{10^{-5} \text{ eV}^2}$ ($1\sigma$) & $7.537^{+0.094}_{-0.10}$ & $7.537^{+0.094}_{-0.10}$ \\
\addlinespace
$\frac{\Delta m^2_{\text{ATM}}}{10^{-3} \text{ eV}^2}$ ($1\sigma$) & $+2.511^{+0.021}_{-0.020}$ & $-2.483^{+0.020}_{-0.021}$ \\
\bottomrule
\end{tabular}
\end{table}

For the sterile-neutrino sector, we adopt the results of the pragmatic $3+1$ global fit to short-baseline neutrino oscillation data presented in Ref.~\cite{Gariazzo:2017fdh}. The marginalized $3\sigma$ intervals are: $|U_{e4}|^2=0.0098 \to 0.031$, $|U_{\mu 4}|^2=0.0060 \to 0.026$, $|U_{\tau 4}|^2 \lesssim 0.039$, with $\Delta m^2_{41} \approx 1.7 \text{ eV}^2$. This fit was constructed to alleviate the well-known tension between appearance and disappearance data by selecting a subset of experimental results that are mutually consistent within the sterile-neutrino hypothesis. We implement these constraints in our framework to restrict the model parameter space by requiring the predicted mixing matrix elements to lie within the experimentally allowed intervals. In particular, our analytical expressions for $U_{e4}$, $U_{\mu4}$, and $U_{\tau4}$ are written in terms of the model parameters. Imposing the above constraints therefore allows us to estimate the viable ranges of these parameters and predict other physical observables, including active-sterile mixing parameters and $CP$-violating phases. 

The $\alpha_{\ell}$ couplings were scattered in the range $(-0.3,0.3)$, since they were expected to be small. The mixing angles of the model, $\varphi_{13}$ and $\varphi_{12}$, were taken in the range $(0,2\pi)$, and the lightest neutrino mass was varied between $(0,0.1)~\text{eV}$. In order to analyse the effects on the symmetry breaking, two scenarios were considered, the first where Max$[\hat \delta,\hat \epsilon]\leq 1.0$, which can be considered as the large breaking scenario, and second Max$[\hat \delta,\hat \epsilon]\leq 0.2$, for the small breaking case. 

\begin{figure}
    \centering
    \includegraphics[width=.32\textwidth]{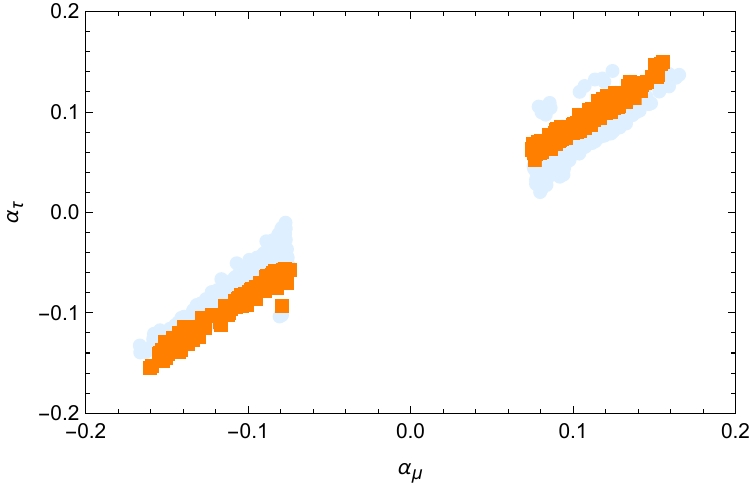}
    \includegraphics[width=.32\textwidth]{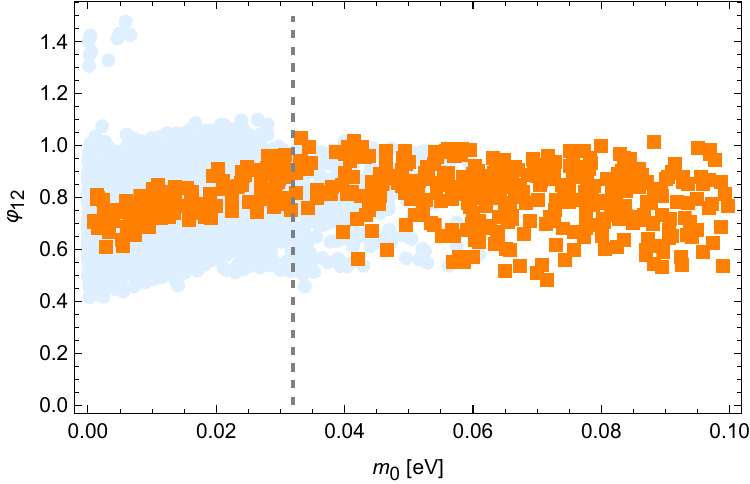}
    \includegraphics[width=.32\textwidth]{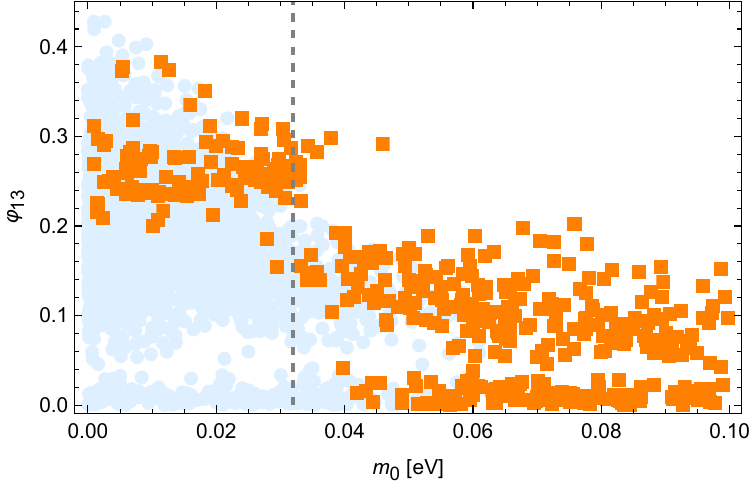} \\
    \includegraphics[width=.32\textwidth]{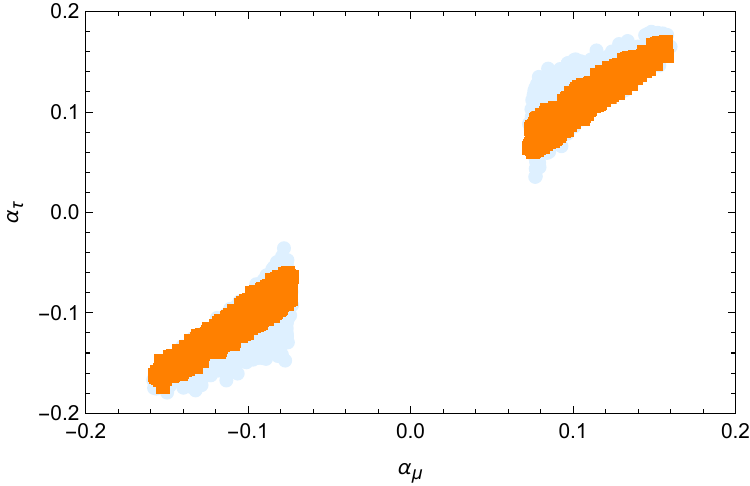}
    \includegraphics[width=.32\textwidth]{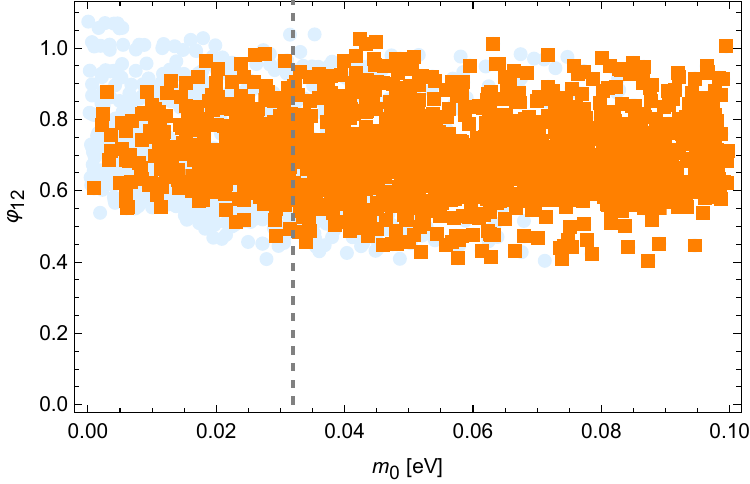}
    \includegraphics[width=.32\textwidth]{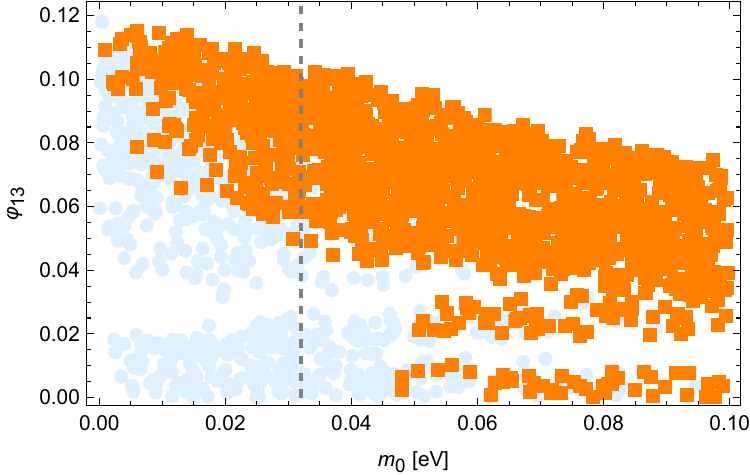}    
    \caption{Allowed regions of model parameters for different breakings scenarios in the normal ordering (upper row) and inverse ordering (lower row). Light blue points corresponds to the case Max$[\hat \delta,\hat \epsilon]\leq 1.0$, and orange squares to Max$[\hat \delta,\hat \epsilon]\leq 0.2$. Vertical dashed line denotes the upper experimental limit on lightest neutrino mass (right hand side is excluded) \cite{Planck:2018vyg,DESI:2024mwx}.}
    \label{fig:paramaters}
\end{figure}

In Fig. \ref{fig:paramaters}, we present the allowed regions of the model parameter space for the two symmetry-breaking scenarios, considering both neutrino mass orderings. The predictions for $\alpha_\ell$ can be understood from Eq. (\ref{eq:column4}). In the regime where the sterile-neutrino mass dominates over the light-neutrino masses, $m_4 \gg m_i$, the latter can be related to the mixing-matrix elements through $U_{\ell 4} = \alpha_\ell + {\cal O}\left(\frac{m_i}{m_4}\right)$. However, Fig. \ref{fig:paramaters} also shows that imposing the small-breaking condition further restricts the allowed coupling parameter space. In this scenario, $\alpha_\mu$ and $\alpha_\tau$ exhibit a clear correlation, with the allowed region favoring parameter combinations with the same sign. This behavior can be inferred from Eq. (\ref{eq:parameters}), since maintaining a small value of $\hat{\delta}$ requires cancellations among these couplings.

On the other hand, the allowed parameter space for the model mixing angles is shown as a function of the lightest neutrino mass, $m_0$. It can be observed that the predicted regions for $\varphi_{12}$ and $\varphi_{13}$ become more constrained once the small-breaking condition, Max$[\hat{\delta},\hat{\epsilon}] \leq 0.2,$ is imposed. In particular, this requirement restricts the allowed values of these mixing angles to well-defined intervals associated with specific ranges of $m_0$. This feature reflects the reduced freedom of the model parameters in the small-breaking regime and, consequently, enhances the correlations among the mixing angles and the neutrino mass scale. For comparison, we also indicate the current upper bound on the lightest neutrino mass inferred from cosmological observations, including the Planck results and recent DESI data~\cite{Planck:2018vyg,DESI:2024mwx}. The cosmological constraint further restricts the viable region of the model parameter space, selecting particular ranges of the mixing angles for small values of $m_0$. Therefore, the combination of the small-breaking condition and experimental limits considerably improves the predictive power of this scenario.

\begin{figure}
    \centering
    \includegraphics[width=.32\textwidth]{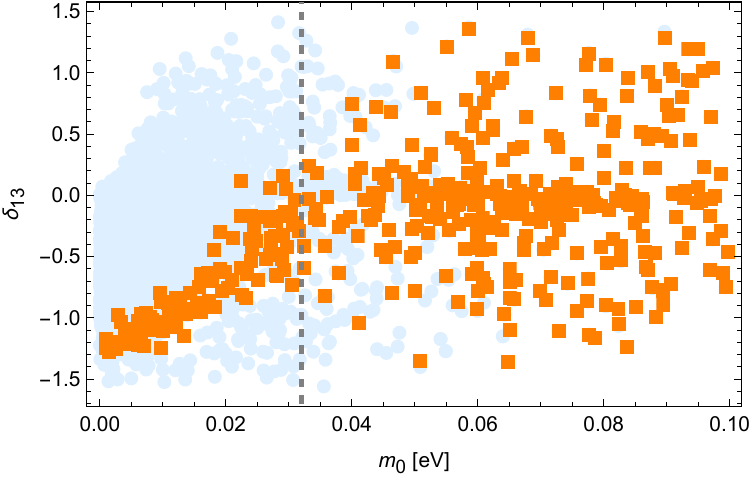}
    \includegraphics[width=.32\textwidth]{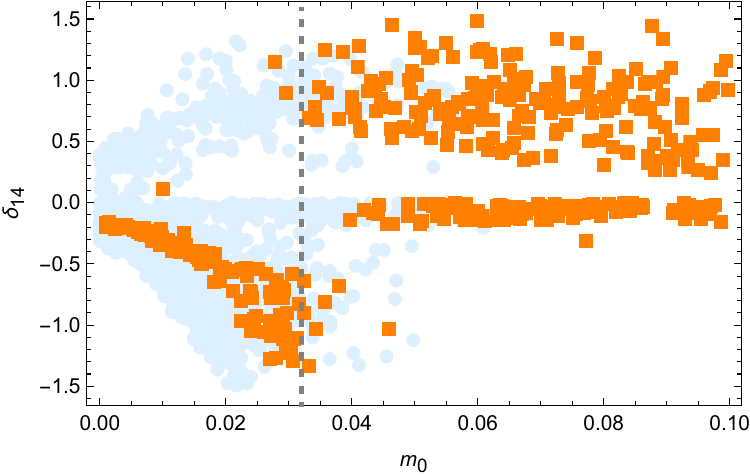}
    \includegraphics[width=.32\textwidth]{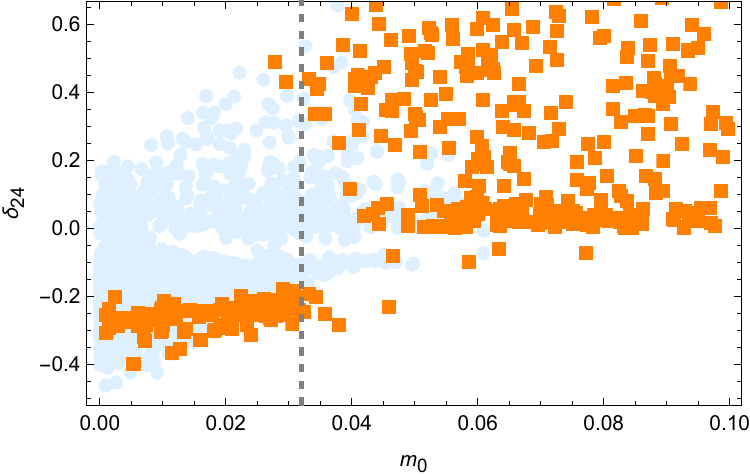} \\
    \includegraphics[width=.32\textwidth]{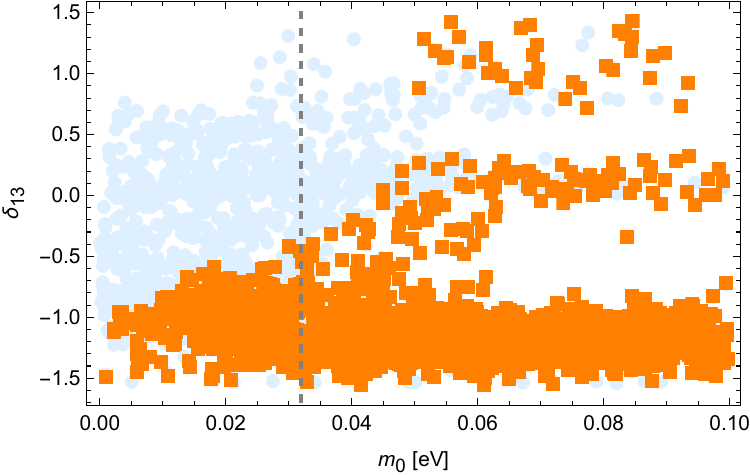}
    \includegraphics[width=.32\textwidth]{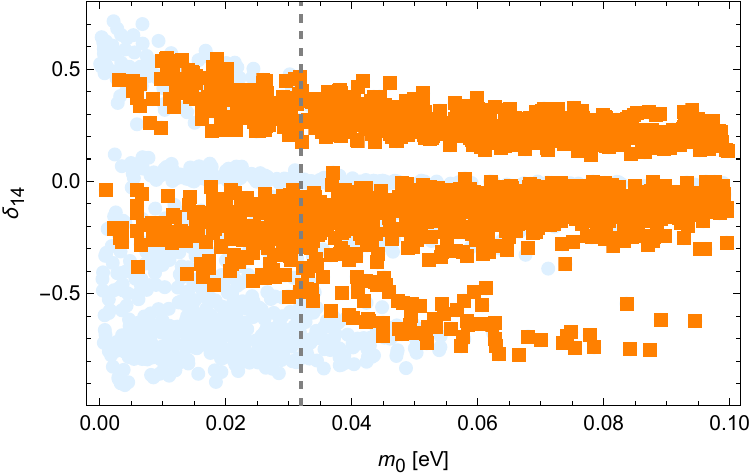}
    \includegraphics[width=.32\textwidth]{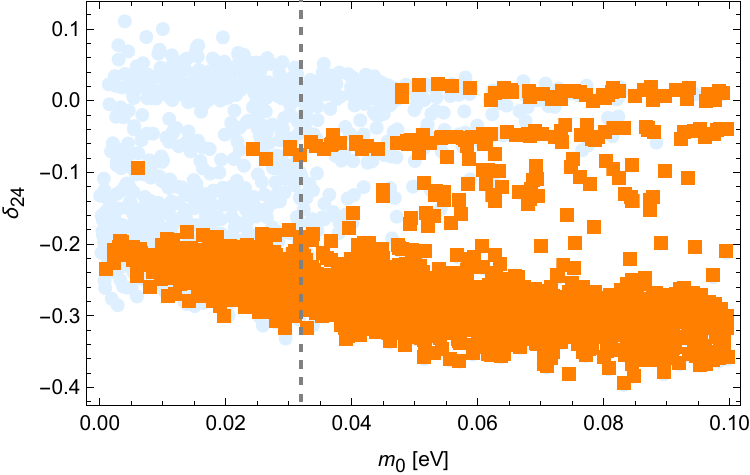}    
    \caption{Allowed regions of Dirac $CP$-violating phases for different braking scenarios. Light blue points corresponds to the case Max$[\hat \delta,\hat \epsilon]\leq 1.0$, and orange squares to Max$[\hat \delta,\hat \epsilon]\leq 0.2$. Vertical dashed line denotes the upper experimental limit on lightest neutrino mass (right hand side is excluded) \cite{Planck:2018vyg,DESI:2024mwx}.}
    \label{fig:dirac}
\end{figure}

Fig. \ref{fig:dirac} displays the predictions for the Dirac CP-violating phases obtained from Eq. (\ref{eq:diracs}), taking into account the global-fit constraints on the active neutrino mixing angles and the active-sterile mixing parameters. A salient feature of the results is that the allowed regions for the $CP$ phases become significantly narrower once the small breaking condition is imposed. This reduction is particularly evident in the low-mass regime, where the lightest neutrino mass remains below the current experimental and cosmological bounds. For $m_0 \lesssim 0.01\,\text{eV}$, the Dirac phases do not span their entire physical ranges, instead, they converge toward specific nonzero intervals in both the NO and IO. We observe in this case that $\delta_{13} \sim (-1.5,1.0)\,\text{rad}$ and $\delta_{24} \sim (-0.4,-0.2)\,\text{rad}$. In contrast, $\delta_{14} \sim -0.2\,\text{rad}$ for normal ordering, whereas in the inverted ordering case it exhibits two approximately degenerate allowed regions around $\delta_{14} \sim -0.1\,\text{rad}$ and $\delta_{14} \sim 0.4\,\text{rad}$. These results show that the small-breaking regime not only restricts the viable parameter space but also generates nontrivial correlations between the absolute neutrino mass scale and the CP-violating sector. In particular, such a nonvanishing values implies that CP violation can remain phenomenologically relevant even for a lightest neutrino mass well below the present upper limits. The emergence of localized intervals for the Dirac phases enhances the predictive power of the small symmetry breaking scenario and distinguishes this realization from more general active-sterile neutrino scenarios, where the CP phases are typically less constrained \cite{Becerra-Garcia:2021gcd}. Consequently, future improvements in the determination of the neutrino mass ordering, the absolute neutrino-mass scale, and the active-sterile mixing parameters may provide a direct test of these predictions and could either further constrain or rule out the small breaking realization of the model.

The Majorana phases appearing in Eq.~(\ref{eq:majoranas}) can only be probed through lepton number violating processes. Nevertheless, the phenomenological implications of different combinations of these phases can be assessed by means of the effective Majorana mass, $|m_{\beta\beta}|$, which governs the rate of neutrinoless double-beta decay  ($0\nu\beta\beta$)~\cite{KamLAND-Zen:2022tow,nEXO:2021ujk}. In the $3+1$ scenario, this quantity is given by
\begin{equation}
|m_{\beta \beta}| = \left|
m_1 U_{e1}^2 + m_2 U_{e2}^2 e^{i\alpha} + m_3 U_{e3}^2 e^{i\beta} + m_4 U_{e4}^2 e^{i\gamma} \right|~,
\end{equation}
where the contribution associated with the sterile state can significantly modify the standard three neutrino expectation, depending on the active-sterile mixing and on the relative values of the Majorana phases.

\begin{figure}
    \centering
    \includegraphics[width=.6\textwidth]{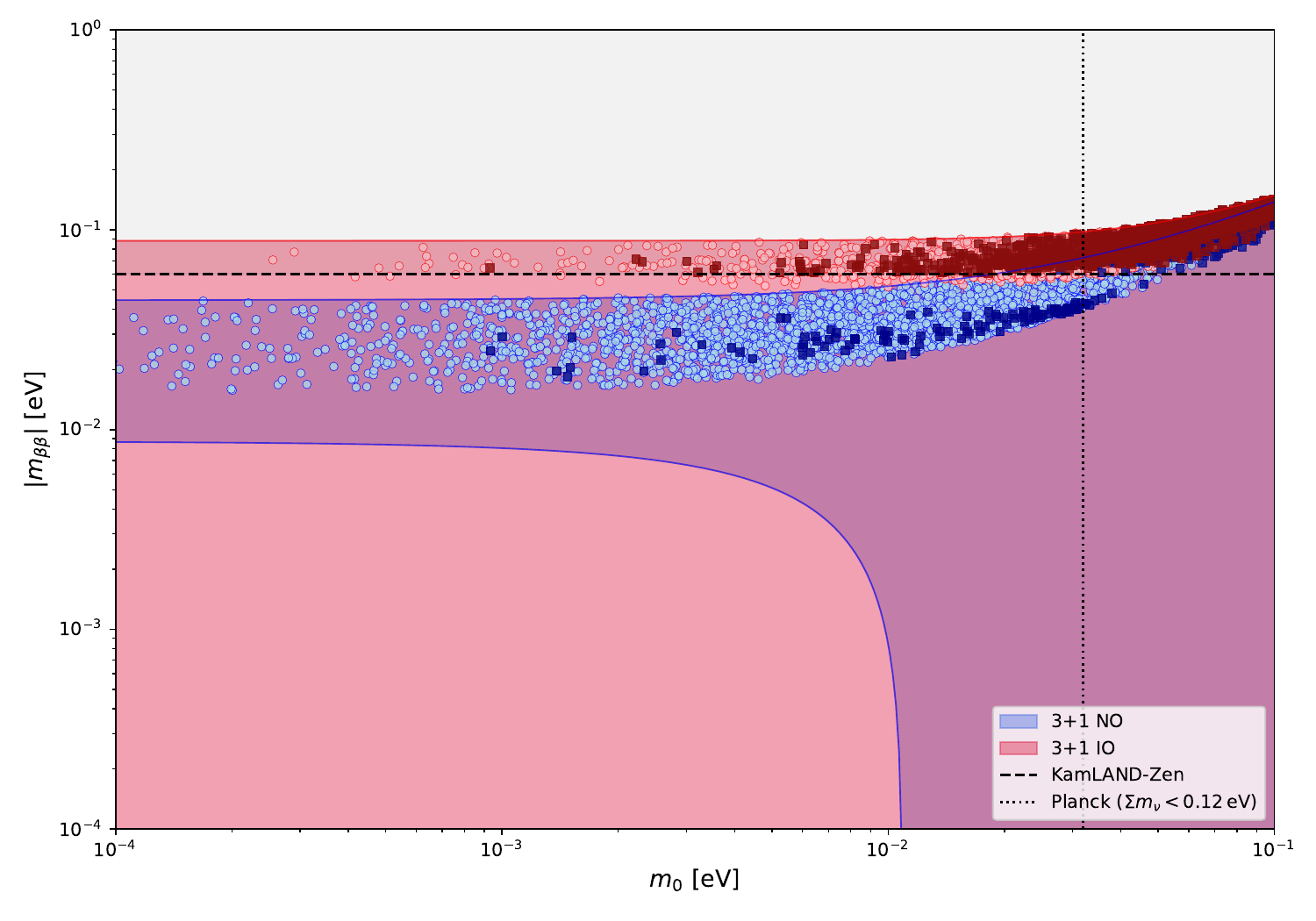}   
    \caption{Region for neutrinoless double beta decay amplitude $m_{\beta \beta}$ in terms of the lightest neutrino mass $m_0$. Light blue (red) points corresponds to the case Max$[\hat \delta,\hat \epsilon]\leq 1.0$, and dark blue (red) squares to Max$[\hat \delta,\hat \epsilon]\leq 0.2$ in the normal (inverse) ordering. Horizontal dashed and vertical dotted lines denotes the current limits from $0\nu\beta\beta$ \cite{KamLAND-Zen:2022tow,nEXO:2021ujk} and cosmology experiments \cite{Planck:2018vyg,DESI:2024mwx}, respectively.}
    \label{fig:doublebeta}
\end{figure}

Figure~\ref{fig:doublebeta} shows the predicted regions for $|m_{\beta\beta}|$ as a function of the lightest neutrino mass, for both neutrino mass orderings. For comparison, we also display the regions obtained by allowing the Majorana CP phases to vary freely, together with the current constraints from $0\nu\beta\beta$ searches~\cite{KamLAND-Zen:2022tow,nEXO:2021ujk} and the upper bound on the lightest neutrino mass inferred from cosmological observations~\cite{Planck:2018vyg,DESI:2024mwx}. The combination of these limits substantially constrains the allowed parameter space, particularly in the IO case. In fact, within the parameter ranges considered here, the small-breaking scenario is excluded for IO, while only a highly restricted region remains viable for the less stringent condition $\max[\hat{\delta},\hat{\epsilon}] \leq 1.0$. This surviving region lies close to the sensitivity reach of next-generation $0\nu\beta\beta$ experiments, implying that future null results could further disfavor, or potentially exclude, this realization of the model in the IO spectrum.

A less restrictive region is obtained for normal ordering (NO). In particular, the small-breaking scenario predicts values as large as $|m_{\beta\beta}| \approx 0.02\,\text{eV}$, which are within the target sensitivity of forthcoming experimental searches. Therefore, an observation of $0\nu\beta\beta$ in this range would provide an important probe of the predicted correlations between the Majorana phases, the active-sterile mixing parameters, and the neutrino mass spectrum. Moreover, Fig.~\ref{fig:doublebeta} shows that, for both mass orderings, the small-breaking condition introduces a nonzero lower limit on the lightest neutrino mass, namely $m_0 \gtrsim 10^{-3}\,\text{eV}$. This lower bound is a distinctive consequence of the restricted phase structure imposed by the small-breaking regime and illustrates that the model does not allow an arbitrarily hierarchical light-neutrino spectrum. Taken together, these results show that $0\nu\beta\beta$ searches and cosmological measurements provide complementary and powerful tests of the model, with the potential to strongly constrain its symmetry breaking structure and CP-violating sector.

\section{Conclusions}
\label{sec:conclusions}

The approximate realization of cobimaximal mixing provides a useful guiding principle for understanding the structure of the lepton sector in the presence of a light sterile neutrino. We have analysed the deviations from the cobimaximal limit by controlling the breaking of the underlying $\mu-\tau$ reflection symmetry induced by the additional sterile state. We have shown that the same parameters that account for deviations from the symmetric limit also determine the size and flavor structure of the active-sterile mixing.

We observed that a small-breaking realization is significantly more predictive than a generic $3+1$ scenario. Requiring the symmetry-breaking parameters to remain small restricts not only the allowed active-sterile mixing strengths, but also their relative signs and correlations. In particular, the correlation between $\alpha_\mu$ and $\alpha_\tau$ reflects the cancellations required by the approximate $\mu-\tau$ reflection structure. Thus, the sterile neutrino sector cannot be regarded as an independent addition to the active neutrino mass matrix, rather, it is directly connected to the mechanism responsible for the departure from cobimaximal mixing.

The approximate symmetry scenario also imposed constrains on the Dirac phases with localized nonzero intervals, especially for low values of the lightest neutrino mass. Therefore, the framework offers a concrete realization in which deviations from the cobimaximal predictions for the atmospheric angle and CP violation can be associated with measurable active-sterile effects. The emergence of correlations between $m_0$, the mixing angles, and the Dirac phases is particularly relevant because it transforms future improvements in oscillation measurements into direct tests of the flavor symmetry breaking pattern.

Predictions for the effective Majorana mass, $|m_{\beta\beta}|$, differ from those obtained by treating the phases as unconstrained. The resulting tension with current neutrinoless double-beta decay limits shows that lepton number violating observables can test the viability of approximate cobimaximal mixing in a way that oscillation data alone cannot. In particular, the exclusion of the small breaking regime for inverted ordering indicates that this spectrum is difficult to reconcile simultaneously with a light sterile state, approximate $\mu-\tau$ reflection symmetry, and present bounds on $0\nu\beta\beta$ decay. By contrast, normal ordering remains a phenomenologically viable and experimentally accessible possibility. The predicted values of $|m_{\beta\beta}|$ can lie within the sensitivity range of forthcoming neutrinoless double beta decay searches, while the lower bound on $m_0$ implied by the small breaking condition prevents an arbitrarily hierarchical light neutrino spectrum. Consequently, a combined analysis of future $0\nu\beta\beta$ results, cosmological constraints on the absolute neutrino mass scale, and improved measurements of active-sterile mixing may decisively test this realization of approximate cobimaximal mixing.


\section*{Acknowdlegments}
The authors acknowledge \textit{Dirección General de Investigaciones} of \textit{Universidad Santiago de Cali} for financial support under grant No. 935-621124-655, and \textit{Universidad de Antioquia} for the facilities to consolidate the final version of this work.

\end{document}